\documentclass[prb,aps,twocolumn,floatfix,a4paper,showpacs]{revtex4} 
\usepackage{epsfig}
\usepackage{umlaut}

\newcommand{\cmta}{Hg$_{\scriptstyle 0.3}$Cd$_{\scriptstyle 0.7}$Te\,}

\DeclareMathAlphabet{\mathitb}{OT1}{cmr}{bx}{sl}

\begin{document}

\title{Giant spin-orbit splitting in a HgTe quantum well}

\author{Y. S. Gui$^{1,2}$}
\author{C. R. Becker$^1$}
\email{becker@physik.uni-wuerzburg.de}
\author{N. Dai$^2$}
\author{J. Liu$^1$}
\author{Z. J. Qiu$^2$}
\author{E.~G.~Novik$^1$}
\author{M.~Sch\"{a}fer$^1$}
\author{X. Z. Shu$^2$}
\author{J. H. Chu$^2$}
\author{H. Buhmann$^1$}
\author{L. W. Molenkamp$^1$}
\affiliation{$^1$Physikalisches Institut der Universit\"at
W\"urzburg, Am  Hubland, 97074 W\"urzburg, Germany}
\affiliation{$^2$National Laboratory for Infrared Physics, 
Shanghai Institute of Technical
Physics,Chinese Academy of Sciences, Shanghai 200083, China}

\date{\today}

\begin{abstract}
We have investigated beating patterns in Shubnikov-de Haas
oscillations for HgTe/Hg$_{0.3}$Cd$_{0.7}$Te(001) 
quantum wells with electron densities of 2 to 3 
~$\times~10^{12}$ cm$^{-2}$. Up to 12 beating nodes have been
observed at magnetic fields between 0.9 and 6 T.
Zero magnetic field spin-orbit splitting energies up to 
$~30$~meV have been directly determined from the node positions as 
well as from the intersection of self-consistently calculated 
Landau levels. These values, which exceed the thermal broadening
of Landau levels, $k_BT$, at room temperature,
are in good agreement with Rashba spin-orbit splitting energies 
calculated by means of an $8~\times~8 \,{\mathitb k} \cdot {\mathitb p}$  
Kane model. The experimental Shubnikov-de Haas 
oscillations are also in good agreement with numerical simulations based on
this model.

\end{abstract}

\pacs{71.70.Ej, 73.21.-b, 72.20.My }

\maketitle

\section{Introduction}

In general, level splitting due to structure inversion asymmetry, 
SIA, known as Rashba {\sc s-o} splitting\cite{Rashba1984,Zawadzki} 
is inversely proportional to the energy gap. {\sc s-o} coupling is 
zero for s-like conduction bands and strong in p-like hole states.
However, mixing of the conduction subbands with the valence subbands 
increases with decreasing energy gap. 
It has been shown that electrons in narrow gap heterostructures based on 
HgTe\cite{Zhang01}, exhibit strong
Rashba {\sc s-o} coupling. In addition to the small energy gap
in HgTe quantum wells, QW's, 
another important factor contributing to the large magnitude of the 
Rashba {\sc s-o} coupling is the inverted band structure of HgTe QW's
with well widths greater than 6~nm, in 
which the first conduction band has heavy hole 
character.\cite{Zhang01,Pfeuffer00}

For possible applications in
spintronics,\cite{Datta90} the Rashba effect has recently been
investigated in a number of narrow gap III-V 
systems\cite{InGaAs1,InGaAs2,InGaAs3,InGaAs4} 
in which typical values of the Rashba  {\sc s-o}  splitting energy, 
$\Delta_R$, are 3 to 5~meV. 
$\Delta_R$ is appreciably larger
in II-VI HgTe QW's, and values of 10 to 17 meV have been 
determined.\cite{Zhang01,Schultz96,Gui04} Zhang {\it et al.}\cite
{Zhang01} demonstrated that the Rashba {\sc s-o} interaction is the
dominant mechanism in such structures; they studied the strong 
dependence of {\sc s-o} splitting on gate voltage and its subsequent 
disappearance when the QW was symmetric as expected for the Rashba effect. 
Compared to the observation of a series of nodes in 
Shubnikov-de Haas, SdH, oscillations for an In$_{1-x}$Ga$_x$As 
heterostructure\cite{Das89}  at $B~<~1$~T, similar beating patterns 
are observable at higher magnetic fields in HgTe 
heterostructures\cite{Zhang01} due to its larger Rashba effect.

In this article, we report on an investigation of beating patterns in
the SdH oscillations in high quality $n$ type HgTe/\cmta{} QW's. 
Up to 12 nodes have been observed in the beating pattern within a 
magnetic field range of 0.9 T $<B<$6 T. A {\sc s-o} splitting 
of $\sim 30$~meV due to the Rashba effect has been
directly deduced from the node positions.
This value is in good agreement
with self-consistent Hartree calculations.  
The observed SdH oscillations and beating patterns are also in good
agreement with the density of states, DOS, obtained from 
self-consistent $\mathitb {k}\cdot \mathitb {p}$ calculations.

\section{Experimental details}

Fully strained $n$ type HgTe/Hg$_{0.3}$Cd$_{0.7}$Te(001) QW's were grown
by molecular beam epitaxy, MBE, on Cd$_{0.96}$Zn$_{0.04}$Te(001)
substrates in a Riber 2300 MBE system. Details of the growth
has been reported elsewhere.\cite{Zhang01,Becker00} Samples A and B 
are from the same chip, Q1772, which was modulation doped asymmetrically in the
top barrier of the HgTe QW structure using CdI$_2$ as a doping
material. The HgTe well width is 12.5 nm and the
Hg$_{0.3}$Cd$_{0.7}$Te barriers consist of a 5.5 nm thick
spacer and a 9 nm thick doped layer. With a well width of 12.5 nm, the
first conduction band in the QW has heavy hole character,  
i.e., is a pure heavy hole state at ${\mathitb k} = 0$,
and following standard nomenclature is labeled $H1$. 

Standard Hall bars were fabricated
by wet chemical etching. A 200 nm thick Al$_2$O$_3$ film was
deposited on top of the structure, which serves as an insulating
layer. Finally Al was evaporated to form a metallic gate
electrode on sample B. A metallic gate was not fabricated on sample A,
which accounts for the different two dimensional electron gas, 2DEG, 
concentrations in these two samples. Ohmic indium contacts to the Hall bars 
were formed by thermal bonding.

Magneto-transport measurements were carried out in several 
different cryostats using
dc techniques with currents of 1 to 5~$\mu$A in magnetic fields
ranging up to 15 T and temperature from 1.4 to 35 K. During the
measurement, the applied electric field was kept low enough
to avoid excessive electron heating.\cite{Gui04b} 

\section{Theoretical details}

The band structure, Landau levels, LL's, and Rashba {\sc s-o} splitting
energy, $\Delta_R$ were obtained from
self-consistent Hartree calculations based on an $8 \times~8
~\mathitb {k}\cdot \mathitb {p}$ band structure model including all
second order terms in the conduction and valence band blocks of
the $8~\times~8$ Hamiltonian. In the calculations 
the inherent inversion asymmetry of  
HgTe and Hg$_{1-x}$Cd$_x$Te has been neglected, because this effect has 
been shown to be very small in narrow gap systems.\cite{Weiler81,Landwehr89}
The envelope function approximation was used to calculate the subbands of
the QW's and the influence of the induced free carriers has been included 
in a self-consistent Hartree calculation. The valence band offset between 
HgTe and CdTe was taken to be 570 meV,\cite{Becker00} and to vary 
linearly with barrier composition.\cite{Shih} 
The band structure parameters of HgTe and CdTe 
at 0 K employed in this investigation are listed in \mbox{Table \ref{table1}} 
and the model is described in detail
elsewhere.\cite{Zhang01,Pfeuffer00}  

The HgTe QW's in this investigation have a well width of 12.5 nm and 
consequently the band structure is inverted. In other words, the
first conduction band in the QW has heavy hole character,  
i.e., is a pure heavy hole state at ${\mathitb k} = 0$,
and following standard nomenclature is labeled $H1$. This has 
important consequences for the large magnitude of the {\sc s-o}
splitting of the $H1-$ and $H1+$ subbands.    

\begin{table}[t]
\begin{minipage}{80mm}
\caption{Band structure parameters employed in the calculations for 
HgTe and CdTe at T=0 K in the $8 \times 8 \,\bf{k\cdot p}$ Kane model}
\begin{center}                                    
\begin{tabular}
{cccccccccc} \hline\hline
  & $E_g$ & $\Delta$ & $E_p$ & $F$ & $\gamma_1$ & $\gamma_2$ 
& $\gamma_3$ & $\kappa$ & $\epsilon$\\ \hline
& eV & eV & eV & &&&&&\\ \hline 
HgTe & -0.303  & 1.08 & 18.8 & 0 & 4.1 & 0.5 & 1.3 & -0.4 & 21\\ \hline
CdTe & 1.606   & 0.91 & 18.8 & -0.09 & 1.47 & -0.28 & 0.03 & -1.31 & 10.4\\ 
\hline \hline
\end{tabular}
\label{table1}
\end{center}
\end{minipage}
\end{table}

\section{Results and Discussion}

\begin{figure} [t]
\centerline{{\epsfig{figure=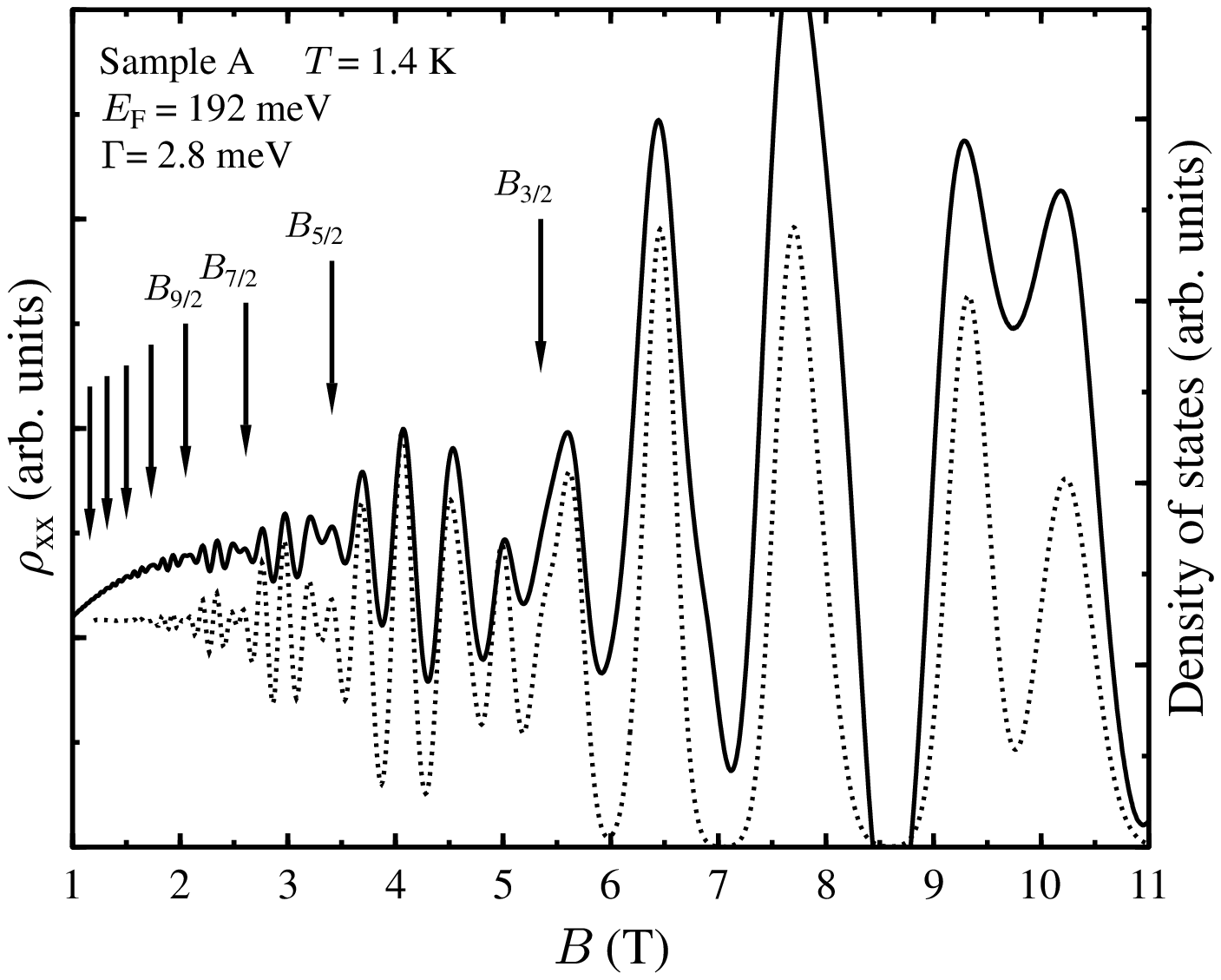,width=84mm}}} 
\caption{SdH oscillations (solid curve) and calculated density of states 
(dotted curve) for sample A. A linear background has been subtracted
from the experimental SdH results. Node
positions in the beating patterns are indicated with
arrows.} 
\label{q1772a-rxxdos}
\end{figure}

\begin{figure} [t]
\centerline{{\epsfig{figure=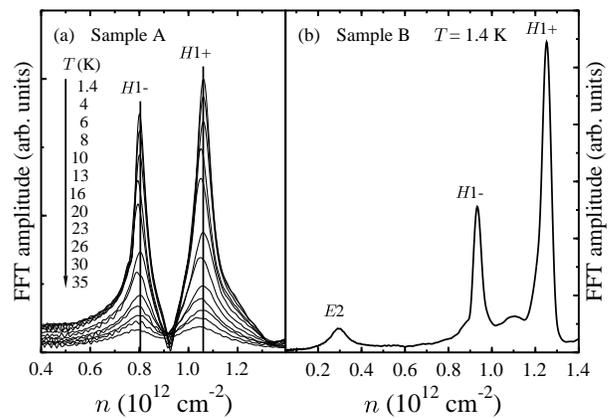,width=84mm}}} 
\caption{(a) FFT of SdH oscillations of sample A for temperatures
between 1.4 and 35~K. The vertical lines are merely guides to the
eye indicating the electron concentrations for the $H1-$ and $H1+$
subbands. (b)  FFT of SdH oscillations of sample B at
1.4~K.}
\label{q1772ab-fft}
\end{figure}

Typical SdH oscillations 
are shown in Fig.~\ref{q1772a-rxxdos} for sample A with a Hall concentration of
$2.0~\times~10^{12}$ cm$^{-2}$ and a mobility of $9.5~\times~
10^4$ cm$^{2}$/Vs at 1.4 K. Oscillations can be resolved
down to 0.8~T, indicating the excellent quality of the sample.
Fast Fourier transformation, FFT, spectra of SdH oscillations are shown 
in Fig.~\ref{q1772ab-fft}(a) 
at various temperatures for sample A.
The 2DEG concentrations of the {\sc s-o} split
$H1-$ and $H1+$ subbands, are 0.80 and $1.06~\times~10^{12}$~cm$^{-2}$,
respectively, which are constant, within experimental uncertainties,
for temperatures up to at least 35 K.   
The amplitudes of the two peaks have  similar temperature
behavior which can be described by\cite{Coleridge} 
\begin{equation}
A(T) = \frac{X}{sinh(X)}
\end{equation}
where
\begin{equation}
X = 2\pi^2 \frac{k_B T}{\hbar \omega_c}
\end{equation}
From the temperature
dependence of the SdH oscillation amplitudes, the effective electron 
mass at the Fermi level, $m_{\rm F}$, was deduced to be 
$0.044 \pm 0.005~m_{e}$ and $0.050 \pm 0.005~m_{e}$ for samples A and B,
respectively, where $m_e$ is the free electron mass. These values 
are in good agreement with calculated effective electron
masses of 0.049 and $0.053~m_e$, respectively. 

Beating patterns in the SdH oscillations are observed when $B~>~0.9$ T.
In the presence of significant broadening of the LL's, the amplitude
of the beat frequency will have a maximum in the vicinity of the 
intersection of two LL's. A node between two maxima will appear 
where only one LL is present i.e.,
$\delta\slash\hbar\omega_c = (N + 1/2)$ with $N=0, 1, 2\ldots$,
where $\delta$ is the total spin splitting and $\hbar\omega_c$ is
the Landau level splitting.\cite{Teran02} The three observable quantum Hall 
plateaus directly below the node at 5.35~T correspond to even filling factors, 
whereas the three above correspond to odd filling factors. 
This node is due to the crossing point at $\delta = 3/2\hbar\omega_c$.

\begin{figure} [t]
\centerline{{\epsfig{figure=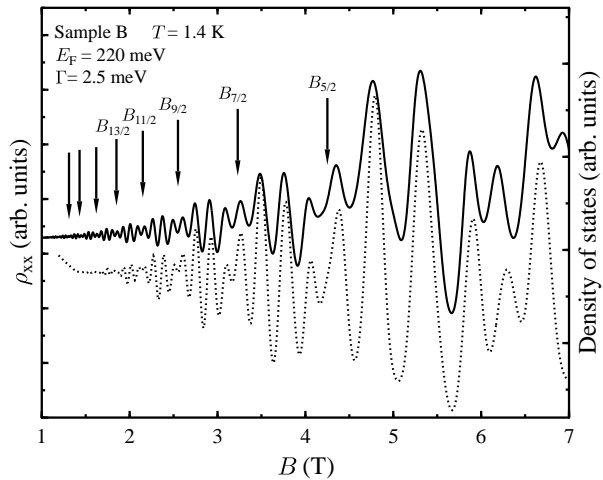,width=84mm}}} 
\caption{SdH oscillations (solid curve) and calculated density of states 
(dotted curve) for sample B. A linear background has been subtracted
from the experimental SdH results. Node
positions in the beating patterns are indicated with
arrows.}           
\label{q1772b-rxxdos}
\end{figure}

Sample B has a higher electron concentration due to deposition of an
insulating layer and metallic gate electrode which results in a
different work function between the semiconductor and surface. 
In Fig.~\ref{q1772b-rxxdos} the vertical arrows 
indicate the node positions of the 
SdH oscillations. Total electron concentration from the 
FFT of $2.76~\times~10^{12}$ cm$^{-2}$ 
($n_{H1+}+n_{H1-}+2n_{E2}$), shown in
Fig.~\ref{q1772ab-fft}(b) agrees well with the value 
of $2.7~\times~10^{12}$ cm$^{-2}$ deduced from the low
magnetic field Hall coefficient. From the ratio of the magnetic field
strengths of all observed nodes, it has been determined that the node
at 4.25 T for sample B corresponds to $\delta = 5/2\hbar\omega_c$. 
Up to 12 beating nodes were observed in the
magnetic field between 0.9~T and the highest field of 7.0~T. 
The second conduction subband, $E2$, is also occupied; however, the
expected weak splitting of this primarily $s$-like state
of $\le 0.2$~meV was less than the experimental resolution.

\begin{figure} [t]
\centerline{{\epsfig{figure=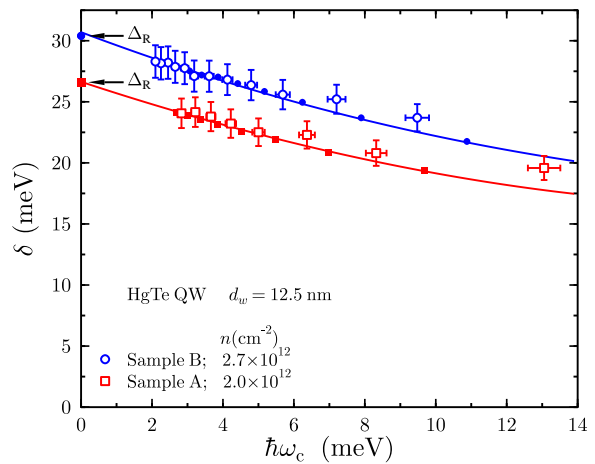,width=84mm}}} 
\caption{Total experimental (open symbols) spin splitting energies and 
values calculated from the intersection of LL's (filled symbols) 
for samples A and B as a
function of $\hbar\omega_c $. The numerically calculated 
Rashba {\sc s-o} splitting energies at $B = 0$ (filled symbols), $\Delta_R$
are indicated by horizonal arrows. 
The lines are least square fits of the analysis 
of the self-consistently calculated LL's described in the text.} 
\label{delta-hwc}
\end{figure}

\begin{figure} [t]
\centerline{{\epsfig{figure=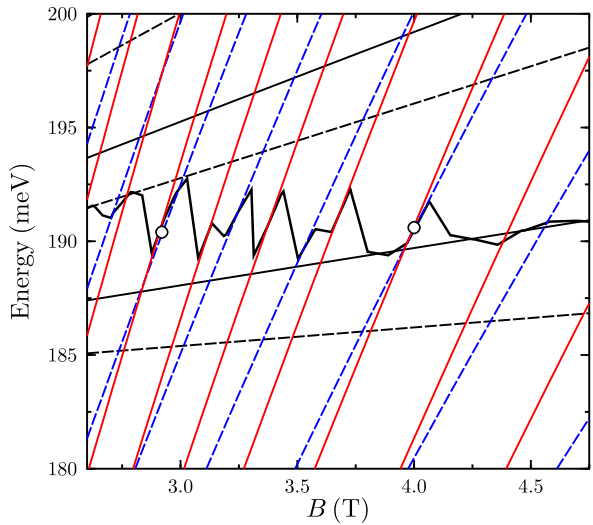,width=84mm}}} 
\caption{Landau levels, LL's, for sample A between $B = 2.6$ and 
4.75 T near the chemical potential, which is reproduced as a thick line.
The nearly vertical lines are LL's of the $H1$ conduction subband.
The intersection of two LL's from the $H1$ 
conduction subband at the chemical potential are indicated with a circle. 
The nearly horizontal lines are LL's of the $E2$
conduction subband.}
\label{ll-q1772a}
\end{figure}

The total spin splitting energy, $\delta$, 
deduced from the node position 
in the beating patterns of the SdH oscillations is shown in 
Fig.~\ref{delta-hwc} as a function of Landau splitting energy, $\hbar\omega_c$.
When the LL's from the $H1$ subband intersect at or near the 
chemical potential 
as shown for sample A in Fig.~\ref{ll-q1772a}, a maximum in the 
amplitude of the beat frequency occurs. $\delta$ can be determined
from the intersection according to    
\begin{eqnarray}
E_{n_i}^- & = & E_{n_f}^+\\
(n_i+1/2)\hbar\omega_c + \delta & = & (n_f+1/2)\hbar\omega_c \\
\label{total}
 \delta & = & \Delta n\cdot \hbar\omega_c
\end{eqnarray}

The two crossing points in Fig.~\ref{ll-q1772a} correspond to a 
$\Delta n$ of 2 and 3.
The change in Landau quantum number for all pairs of LL's which 
intersect at or near the chemical potential
is given by 1, 2, 3, 4, 5 $\dots$. 
Even though the LL's with $\Delta n = 1$ do intersect, they do so 
further removed from the chemical potential than the subsequent 
series of LL pairs. 
In order to increase the number of theoretical data points, the energy 
difference between appropriate LL's was employed when one LL was below 
the chemical potential and the other above. These values are in excellent 
agreement with those obtained from the intersection of LL's. 
A similar series of LL's crossing points exist for sample B, and
the analysis of the LL's in the vicinity of the chemical potential
described above also resulted in a consistent set of data.

Values of $\delta$ obtained from the intersection of LL's 
and a least square fit all theoretical data for both samples are 
plotted as a function of $\hbar\omega_c$ in Fig.~\ref{delta-hwc}
together with the experimental results. Obviously 
theory and experiment are in very good agreement with the exception 
of the $B_{3/2}$ node in the amplitude of the beat frequency for
sample A, which corresponds to LL's with small filling factors.

\begin{figure}[t]
\centerline{{\epsfig{figure=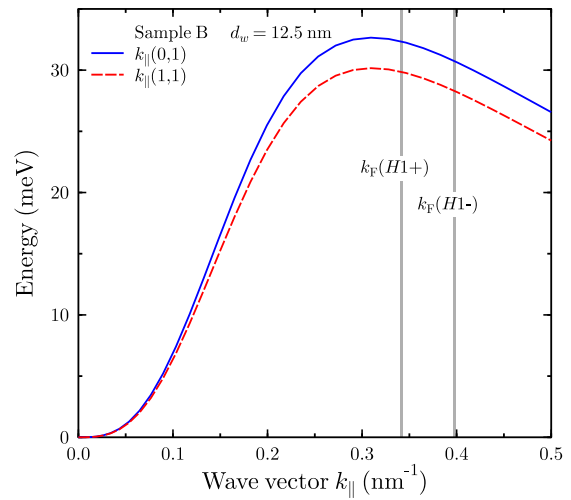,width=80mm}}} 
 \caption{Calculated spin splitting energy of the $H1$ subbands
for sample B. The ${\mathitb k}$ vector for the $H1-$ and $H1+$ subbands   
at the Fermi surface are indicated by vertical lines.}
\label{delta-ek0}
\end{figure}

The calculated Rashba {\sc s-o} splitting energies, $\Delta_R$, for
sample B are 31.5 and 29.1~meV for the in-plane 
$k_\|(0,1)$ and $k_\|(1,1)$ vectors at the Fermi surface, 
respectively, see Fig.~\ref{delta-ek0}. Similarly
the values for sample A are 27.5 and 25.4~meV, respectively. 
$\Delta_R$ values averaged over $\mathitb k_{\parallel}$ space
of 26.5 and 30.4 meV for samples A and B are in good agreement with
the experimentally determined total {\sc s-o} splitting energies
of $26 \pm 1$ and $30 \pm 1$~meV, respectively. 
The experimental splitting is due to the Rashba {\sc s-o} effect,
which results in the large population difference of 14
and 14.7~\% for samples A and B, respectively, shown in the FFT
spectra in Fig.~\ref{q1772ab-fft}.

The Rashba {\sc s-o} splitting energy of up to $30$~meV in these 
HgTe QW's is almost one order of magnitude larger 
than the previously reported values in III-V materials.
This is due to the unique band structure of the HgTe system and in 
particular the inverted band structure. This value is also larger than 
previously reported values by Schultz {\it et al.} ~\cite{Schultz96} and 
Zhang {\it et al.}\cite{Zhang01} for HgTe based QW's. 
This is mainly due to a larger 2DEG concentration in the $H1$ subband 
in the present QW's and the larger structure inversion asymmetry.

In order to compare the results of our self-consistent Hartree
calculations with the measured longitudinal resistance, we
have employed the following relationship to calculate the density of states,
DOS, from the Landau level structure in the lowest order cumulant
approximation according to Gerhardts;\cite{Gerhardts76}
\begin{equation}
\label{eq4} D(\varepsilon_n^\pm) = \frac{1}{2\pi
\lambda_c^2}\sum\limits_{n\pm} {\left[ {\frac{\pi }{2}\Gamma _n^2
}\right]^{-1/2} \exp
\left[{-2\frac{(E_F-\varepsilon_n^\pm)^2}{\Gamma _n^2 }}\right]}
\end{equation}
Here $\varepsilon_n^\pm$ are the Landau level energies which
are the result of our self-consistent Hartree calculations.
${\lambda_c=\sqrt{\hbar/eB}}$ is the usual magnetic
length, and $\Gamma_n$ is the Landau level broadening and assumed to
be a constant.

The experimental SdH oscillations and the numerical simulations of the DOS
by means of Eq.~\ref{eq4} for sample A and B are reproduced in 
Figs.~\ref{q1772a-rxxdos} and \ref{q1772b-rxxdos}, respectively. 
The calculated Fermi energy 
was modified less than 1~\% in order to align the SdH oscillations. 
The best fit was obtained using $\Gamma = 2.8$ meV for sample A, 
and $\Gamma = 2.5$~meV for sample B.

\section{Conclusions}

In conclusion, the beating patterns in the SdH oscillations of
modulation doped HgTe/Hg$_{0.3}$Cd$_{0.7}$Te QW's have been
analyzed. The {\sc s-o} splitting energy which has been directly determined 
from the node positions to be as high as 30 meV, is almost one
magnitude higher than that in InGaAs heterostructures with similar
carrier densities. Self-consistent Hartree calculations based on an
$8~\times~8~{\mathitb k}\cdot{\mathitb p}$ Hamiltonian have 
demonstrated that the experimental zero field splitting energies are due to
Rashba {\sc s-o} splitting. Furthermore good agreement between experimental
SdH oscillations and calculated DOS is evidence that the Rashba term is the
dominant mechanism  of giant {\sc s-o} splitting in HgTe QW's with an inverted
band structure. This large $\Delta_R$ in HgTe QW's with an inverted 
band structure is caused by its narrow gap, the large
spin-orbit gap between the bulk valence bands $\Gamma _8^v$ and
$\Gamma_7^v$ and the heavy hole character of the first conduction
subband. Furthermore, our calculations show that the method of directly
deducing {\sc s-o} splitting from node positions in SdH oscillations,
is applicable even for a system with a strongly nonparabolic band structure.

\section{Acknowledgments}

The financial support of the major state basic research project
no. G001CB3095 of China, National Natural Science Foundation of
China (no. 10374094), the Deutsche Forschungsgemeinschaft (SFB410),
BMBF, and the DARPA SpinS program are gratefully acknowledged.


\end{document}